# On optimal filtering of measured Mueller matrices

## José J. Gil

*Universidad de Zaragoza. Pedro Cerbuna 12, 50009 Zaragoza, Spain*
*ppgil@unizar.es*

While any two-dimensional mixed state of polarization of light can be represented by a combination of a pure state and a fully random state, any Mueller matrix can be represented by a convex combination of a pure component and three additional components whose randomness is scaled in a proper and objective way. Such characteristic decomposition constitutes the appropriate framework for the characterization of the polarimetric randomness of the system represented by a given Mueller matrix, and provides criteria for the optimal filtering of noise in experimental polarimetry.

## 1. Introduction

Mueller polarimetry is a powerful tool with growing impact in a great variety of fields like science, industry, medicine and remote sensing. Since experimental polarimetric setups have a limited precision, a given measured Mueller matrix $\mathbf{M}_{\exp}$ is affected by errors, and appropriate procedures for its filtering are required [1-7]. Moreover, the analysis and interpretation of the information contained $\mathbf{M}_{\exp}$ usually requires that the operator takes advantage of his a priori knowledge of the characteristics of the sample. For instance, frequently the operator knows that a certain sample is essentially nondepolarizing so that he is interested in applying appropriate filtering procedures for obtaining the optimal nondepolarizing estimate $\mathbf{M}_J$ of $\mathbf{M}_{\exp}$. In other cases, it is expected that the polarimetric behavior of the sample (for instance, a nondepolarizing element on a homogeneous substrate) is due to two or more independent contributions, so that $\mathbf{M}_{\exp}$ should be filtered to obtain the optimal estimate $\mathbf{M}_f$ representing the properties of the whole sample.

In general, the filtering procedures used in polarimetry are based on the *spectral decomposition* of a depolarizing Mueller matrix $\mathbf{M}$ into four pure components associated respectively with the eigenvectors of the covariance matrix $\mathbf{H}$ associated with $\mathbf{M}$ [8-9]. From the consideration of the recent approaches for the *arbitrary* [10,11] and *characteristic* (or *trivial*) parallel decompositions of $\mathbf{M}$ [11], the filtering criteria are revisited in this work with the aim of obtaining a rigorous characterization of the nature and relative weight of the estimates of $\mathbf{M}$ through the subtraction of the random background. To do so, Section 2 is devoted to the formulation of the arbitrary, spectral and characteristic decompositions of $\mathbf{M}$, which constitute the theoretical framework for the present study; Section 3 deals with the notion of polarimetric randomness, its characterization and interpretation; the covariance filtering is revisited in Section 4 as a first step of the whole filtering process; Section 5 describes an objective procedure for filtering the polarimetric noise; Section 6 is devoted to the procedure for checking if an hypothesized constituent is physically realizable as a parallel component of the measured sample and, if so, to the calculation of its relative weight (concentration, or relative cross section), and Section 7 deals with the analysis and interpretation of the characteristic nondepolarizing (or *pure*) component, which is associated with the largest possible relative weight of a pure component embedded into the sample, thus corresponding to the optimal pure estimate of the sample.

## 2. Parallel decompositions of Mueller matrices

From a polarimetric point of view, any macroscopic linear interaction of polarized light with a material medium can be considered as a statistical result of a number of interactions with elementary components of the medium that behave in deterministic manner and, thus, are nondepolarizing (or polarimetrically *pure*) in the sense that they do not reduce the degree of polarization of the corresponding totally polarized input pencils. Therefore, any Mueller matrix can be expressed as a convex sum of pure Mueller matrices $\mathbf{M}_{Ji}$ (hereafter, when appropriate, the subscript $J$ will be used to distinguish pure Mueller matrices from general ones) [12-14]

$$\mathbf{M} = \sum_{i=1}^{n} p_i \mathbf{M}_{Ji} \quad \left(0 \le p_i,\; \sum_{i=1}^{n} p_i = 1\right) \tag{1}$$

As a consequence of this property, any Mueller matrix $\mathbf{M}$ has an associated positive semidefinite *coherency matrix* $\mathbf{H}(\mathbf{M})$ defined as [1,9]

$$\mathbf{H}(\mathbf{M}) = \frac{1}{4}\begin{pmatrix} m_{00}+m_{01} \\ +m_{10}+m_{11} & m_{02}+m_{12} \\ +i(m_{03}+m_{13}) & m_{20}+m_{21} \\ -i(m_{30}+m_{31}) & m_{22}+m_{33} \\ +i(m_{23}-m_{32}) \\ m_{02}+m_{12} \\ -i(m_{03}+m_{13}) & m_{00}-m_{01} \\ +m_{10}-m_{11} & m_{22}-m_{33} \\ -i(m_{23}+m_{32}) & m_{20}-m_{21} \\ -i(m_{30}-m_{31}) \\ m_{20}+m_{21} \\ +i(m_{30}+m_{31}) & m_{22}-m_{33} \\ +i(m_{23}+m_{32}) & m_{00}+m_{01} \\ -m_{10}-m_{11} & m_{02}-m_{12} \\ +i(m_{03}-m_{13}) \\ m_{22}+m_{33} \\ -i(m_{23}-m_{32}) & m_{20}-m_{21} \\ +i(m_{30}-m_{31}) & m_{02}-m_{12} \\ -i(m_{03}-m_{13}) & m_{00}-m_{01} \\ -m_{10}+m_{11} \end{pmatrix} \tag{2}$$

Thus, a 4×4 real matrix satisfies the *covariance criterion* (or *Cloude's criterion*) expressed in Eq. (1) if and only if the four eigenvalues of $\mathbf{H}(\mathbf{M})$ are nonnegative.

which can always be expressed as

$$\mathbf{H} = m_{00}\, \mathbf{U}\, \mathrm{diag}\!\left(\hat{\lambda}_0, \hat{\lambda}_1, \hat{\lambda}_2, \hat{\lambda}_3\right) \mathbf{U}^{\dagger} \tag{3}$$

where $\mathbf{U}$ is unitary and $\hat{\lambda}_i$ are the (nonnegative) normalized eigenvalues of $\mathbf{H}$

$$\hat{\lambda}_i \equiv \lambda_i / m_{00} = \lambda_i / \mathrm{tr}\,\mathbf{H} \qquad (4)$$

taken in decreasing order $\hat{\lambda}_0 \geq \hat{\lambda}_1 \geq \hat{\lambda}_2 \geq \hat{\lambda}_3$.

Thus, a 4×4 real matrix satisfies the *covariance criterion* (or *Cloude's criterion*) expressed in Eq. (1) if and only if the four eigenvalues of $\mathbf{H}(\mathbf{M})$ are nonnegative.

A complementary physical restriction is that, leaving aside certain artificial arrangements in which the intensity is amplified for certain input polarization states [15], the linear interactions never increase the intensity of the interacting electromagnetic wave, and consequently this *passivity criterion* implies that all the pure Mueller matrices $\mathbf{M}_{Ji}$ entering in Eq. (1) are passive. Let us recall that a necessary and sufficient condition for a pure Mueller matrix $\mathbf{M}_J$ to be passive is [16,17]

$$m_{00}(1+D) \leq 1 \qquad (5)$$

where the upper-left element $m_{00}$ of $\mathbf{M}_J$ is the mean intensity coefficient and $D$ is the diattenuation. Since any physically realizable Mueller matrix can be considered an ensemble average of pure and passive Mueller matrices, the combination of both covariance and passivity criteria characterizing Mueller matrices can be referred to as the *ensemble criterion*.

Let us consider the integer parameter $r \equiv \mathrm{rank}\,\mathbf{H}$ and recall that it has been demonstrated that the covariance matrix $\mathbf{H}$ can always be submitted to its *arbitrary decomposition*, formulated as [11]

$$\mathbf{H} = \sum_{i=0}^{r-1} p_i \left(m_{00}\hat{\mathbf{H}}_{Ji}\right)$$

$$\left[\sum_{i=0}^{3} p_i = 1 \quad \hat{\mathbf{H}}_{Ji} = \mathbf{w}_i \otimes \mathbf{w}_i^\dagger \quad |\mathbf{w}_i| = 1\right] \qquad (6)$$

where the dagger indicates conjugate transpose, $\otimes$ stands for the Kronecker product, $\hat{\mathbf{H}}_{Ji}$ are the normalized *pure covariance matrices* (i.e., with a unique eigenvector with nonzero eigenvalue) generated by respective unit *covariance vectors* $\mathbf{w}_i$ belonging to the $r$-dimensional image subspace of $\mathbf{H}$ (commonly denoted as $\mathrm{Range}\,\mathbf{H}$). The nonnegative coefficients $p_i$ (concentrations or relative cross sections) of the components of the arbitrary decomposition are given by [11]

$$p_i = \frac{1}{\sum_{j=0}^{r-1} \frac{1}{\hat{\lambda}_j} \left|\left(\mathbf{U}^\dagger \mathbf{w}_i\right)_j\right|^2} \qquad (7)$$

In the particular case that the covariance vectors $\mathbf{w}_i$ are taken as being the $r$ eigenvectors $\mathbf{u}_i$ of $\mathbf{H}$ with nonzero eigenvalue, the arbitrary decomposition becomes the *spectral decomposition*

$$\mathbf{H} = \sum_{i=0}^{r-1} \hat{\lambda}_i \left[m_{00}\left(\mathbf{u}_i \otimes \mathbf{u}_i^\dagger\right)\right] \qquad (8)$$

where $\hat{\lambda}_i$ play the role of the relative weights of the spectral components $\mathbf{H}_{Ui} \equiv m_{00}\left(\mathbf{u}_i \otimes \mathbf{u}_i^\dagger\right)$ generated by the respective eigenvectors $\mathbf{u}_i$.

It should be stressed that, although the spectral decomposition is a particular option among the infinite possible arbitrary decompositions, it has the peculiarity that $\hat{\lambda}_{r-1} \leq p_{r-1}$, and $\hat{\lambda}_0 \geq p_0$, where $p_{r-1}$ and $p_0$ are respectively the smallest (but nonzero) and the largest coefficients of the arbitrary decomposition considered (note that the equalities hold only for the spectral decomposition).

While the arbitrary decomposition (and hence the spectral decomposition) is formulated in terms of pure components, $\mathbf{H}$ can also be expressed as convex sums where the addends are not necessarily pure. In particular, due to its key role played for the inspection of the structure of polarimetric randomness of $\mathbf{H}$ (and hence of the associated Mueller matrix $\mathbf{M}$), let us bring up the characteristic (or trivial) decomposition [18,19]

$$\mathbf{H} = P_1 \mathbf{H}_{U0} + (P_2 - P_1)\mathbf{H}_2 + (P_3 - P_2)\mathbf{H}_3 + (1 - P_3)\mathbf{H}_{\Delta 0} \qquad (9)$$

where the covariance matrices of the components are given by

$$\mathbf{H}_{U0} \equiv m_{00}\left[\mathbf{U}\,\mathrm{diag}(1,0,0,0)\mathbf{U}^\dagger\right] = (\mathrm{tr}\,\mathbf{H})\left(\mathbf{u}_0 \otimes \mathbf{u}_0^\dagger\right)$$

$$\mathbf{H}_2 \equiv \frac{1}{2}m_{00}\left[\mathbf{U}\,\mathrm{diag}(1,1,0,0)\mathbf{U}^\dagger\right] = \frac{1}{2}(\mathrm{tr}\,\mathbf{H})\sum_{i=0}^{1}\mathbf{u}_i \otimes \mathbf{u}_i^\dagger$$

$$\mathbf{H}_3 \equiv \frac{1}{3}m_{00}\left[\mathbf{U}\,\mathrm{diag}(1,1,1,0)\mathbf{U}^\dagger\right] = \frac{1}{3}(\mathrm{tr}\,\mathbf{H})\sum_{i=0}^{2}\mathbf{u}_i \otimes \mathbf{u}_i^\dagger$$

$$\mathbf{H}_{\Delta 0} \equiv \frac{1}{4}m_{00}\mathbf{I}_4 = \frac{1}{4}(\mathrm{tr}\,\mathbf{H})\sum_{i=0}^{3}\mathbf{u}_i \otimes \mathbf{u}_i^\dagger \qquad (10)$$

where $\mathbf{I}_4$ is the 4×4 identity matrix, and the nonnegative parameters $P_i$ $(i = 1, 2, 3)$ determining the coefficients of the components are the *indices of polarimetric purity* (IPP) of $\mathbf{M}$ defined as [20]

$$P_1 \equiv \hat{\lambda}_0 - \hat{\lambda}_1$$
$$P_2 \equiv \hat{\lambda}_0 + \hat{\lambda}_1 - 2\hat{\lambda}_2$$
$$P_3 \equiv \hat{\lambda}_0 + \hat{\lambda}_1 + \hat{\lambda}_2 - 3\hat{\lambda}_3 \qquad (11)$$

which satisfy the property

$$0 \leq P_1 \leq P_2 \leq P_3 \leq 1 \qquad (12)$$

Due to the biunivocal relation between $\mathbf{H}$ and $\mathbf{M}$, all the above decompositions formulated in terms of covariance matrices, have their respective counterparts in terms of Mueller matrices. The arbitrary decomposition of $\mathbf{M}$ is formulated as

$$\mathbf{M} = \sum_{i=0}^{r-1} p_i\left(m_{00}\hat{\mathbf{M}}_{Ji}\right) \qquad (13)$$

where $\hat{\mathbf{M}}_{Ji}$ are the normalized pure Mueller matrices associated with the pure covariance matrices $\hat{\mathbf{H}}_{Ji} = \left(\mathbf{w}_i \otimes \mathbf{w}_i^\dagger\right)$, $\mathbf{w}_i$ being a set of $r$ mutually independent four-dimensional unit complex vectors belonging to $\mathrm{Range}\,\mathbf{H}$, while the nonnegative coefficients $p_i$ are determined by Eq. (7).

The spectral decomposition of $\mathbf{M}$ is formulated as

$$\mathbf{M} = \sum_{i=0}^{3} \hat{\lambda}_i \left[m_{00}\hat{\mathbf{M}}_{Ui}\right] \qquad (14)$$

where $m_{00}\hat{\mathbf{M}}_{Ui}$ are the pure Mueller matrices of associated with the covariance matrices $\mathbf{H}_{Ui} \equiv m_{00}\left(\mathbf{u}_i \otimes \mathbf{u}_i^\dagger\right)$ of the spectral components.

The characteristic decomposition of $\mathbf{M}$ is given by

$$\mathbf{M} = P_1 \mathbf{M}_{U0} + (P_2 - P_1)\mathbf{M}_2 + (P_3 - P_2)\mathbf{M}_3 + (1 - P_3)\mathbf{M}_{\Delta 0} \qquad (15)$$

where

- $\mathbf{M}_{U0} \equiv m_{00}\hat{\mathbf{M}}_{U0}$ is the *characteristic component*, whose associated covariance vector is the eigenvector $\mathbf{u}_0$ of $\mathbf{H}$, which in turn corresponds to the largest eigenvalue $\lambda_0$ of $\mathbf{H}$. Thus, $\mathbf{M}_{U0}$ coincides with the first component of the spectral decomposition, while its relative weight with respect to the complete system in the characteristic decomposition is given by $P_1$ (and not by $\hat{\lambda}_0$, as occurs in the spectral decomposition).

- $\mathbf{M}_2 \equiv m_{00} \hat{\mathbf{M}}_2$ is the *2D-depolarizer* constituted by an equiprobable mixture of the first two spectral components, with relative weight $P_2 - P_1$.
- $\mathbf{M}_3 \equiv m_{00} \hat{\mathbf{M}}_3$ is the *3D-depolarizer* constituted by an equiprobable mixture of the first three spectral components, with relative weight $P_3 - P_2$.
- $\mathbf{M}_{\Delta 0} \equiv m_{00} \text{diag}(1,0,0,0)$ represents an *ideal depolarizer*, with relative weight $1 - P_3$.

The above types of parallel decomposition of a general Mueller matrix $\mathbf{M}$ constitute the appropriate framework to address the main aim of this work, consisting of identifying optimal criteria for characterizing the polarimetric randomness (which in certain cases can be interpreted as measurement noise) as well as for filtering measured Mueller matrices.

When dealing with experimentally determined Mueller matrices $\mathbf{M}_{\exp}$ (hence affected by errors) it is very important to note that the coefficients $\hat{\lambda}_i$ of the spectral decomposition of $\mathbf{M}_{\exp}$ are nonnegative inasmuch the IPP of $\mathbf{M}_{\exp}$ are (and vice versa).

## 3. Polarimetric randomness

Polarization algebra involving two-dimensional and three-dimensional states of polarization (determined by the respective 2D and 3D Stokes parameters [21]) as well as the parameters characterizing the polarimetric properties of linear media (Mueller matrix elements) can be formulated in a unified and symmetric treatment based on the key notion of coherency matrix (characterized as being a positive semidefinite Hermitian matrix; that is, with the mathematical structure of a covariance matrix) [22]. In fact, a 2D polarization state can be represented by its 2×2 coherency matrix (also called polarization matrix) $\boldsymbol{\Phi}$; a 3D polarization state can be represented by its 3×3 coherency matrix $\mathbf{R}$ [23], and the polarimetric properties of a linear medium can be represented by means of its associated 4×4 covariance matrix $\mathbf{H}$ (note that an alternative but fully equivalent representation is given by the so-called coherency matrix $\mathbf{C}$ associated with $\mathbf{H}$ and with the corresponding Mueller matrix $\mathbf{M}$ [8,24]).

In the case of 2D states of polarization, it is well-known that $\boldsymbol{\Phi}$ can always be expressed as the convex sum

$$\boldsymbol{\Phi} = \mathcal{P}\,\boldsymbol{\Phi}_P + (1-\mathcal{P})\,\boldsymbol{\Phi}_{u-2D} \qquad (16)$$

where $\mathcal{P}$ is the degree of polarization, $\boldsymbol{\Phi}_P$ represents a pure (or totally polarized) state and $\boldsymbol{\Phi}_{u-2D} = \mathbf{I}_2/2$, where $\mathbf{I}_2$ is the 2×2 identity matrix, represents a unpolarized state (i.e., fully random from the polarimetric point of view).

It is remarkable that such decomposition of the coherency matrix into a pure state and a fully random state is exclusive of two-dimensional systems. In fact, Eq. (16) is not other than the expression of the characteristic decomposition of 2D coherency matrices [11], where the coefficients are fully determined by the single parameter $\mathcal{P}$. Nevertheless, in the case of 3D coherency matrices, there is no parallel decomposition of $\mathbf{R}$ (i.e. a convex sum of 3D coherency matrices that equals $\mathbf{R}$) where all the randomness is allocated into a single component, but the structure of randomness (or, conversely, the structure of polarimetric purity) is determined by the characteristic decomposition [23]

$$I\,\hat{\mathbf{R}} = P_1\,I\,\hat{\mathbf{R}}_P + (P_2 - P_1)\,I\,\hat{\mathbf{R}}_m + (1 - P_2)\,I\,\hat{\mathbf{R}}_{u-3D} \qquad (17)$$

where $I$ is the intensity (power density flux), $(P_1, P_2)$ are the IPP of $\mathbf{R}$, and

$$\hat{\mathbf{R}}_P \equiv \mathbf{U}\,\text{diag}(1,0,0)\,\mathbf{U}^\dagger$$

$$\hat{\mathbf{R}}_m \equiv \frac{1}{2}\mathbf{U}\,\text{diag}(1,1,0)\,\mathbf{U}^\dagger$$

$$\hat{\mathbf{R}}_{u-3D} \equiv \frac{1}{3}\mathbf{U}\,\text{diag}(1,1,1)\,\mathbf{U}^\dagger = \frac{\mathbf{I}_3}{3} \qquad (18)$$

$\mathbf{U}$ being the unitary matrix whose columns $\mathbf{u}_i$ are the eigenvectors of $\mathbf{R}$ and $\mathbf{I}_3$ being the 3×3 identity matrix. $\hat{\mathbf{R}}_P$ represents a normalized pure state, $\hat{\mathbf{R}}_m$ represents an equiprobable mixture of the first two spectral components of $\hat{\mathbf{R}}$, and $\hat{\mathbf{R}}_{u-3D}$ represents a 3D fully random state (3D unpolarized state). Thus, the structure of polarimetric randomness is regulated by the pair of IPP $(P_1, P_2)$, and not by a single overall parameter, like for instance, the 3D degree of polarimetric purity [22,25]

$$P_{3D} = \sqrt{\frac{3\,\text{tr}(\mathbf{R}^\dagger \mathbf{R})}{2\,\text{tr}(\mathbf{R}^2)} - \frac{1}{2}} = \frac{1}{2}\sqrt{3P_1^2 + P_2^2} \qquad (19)$$

Thus, as occurs with $\mathbf{R}$, the polarimetric randomness associated with a given Mueller matrix $\mathbf{M}$ cannot be allocated into a single component, but is shared, in a scaled and objective way, among the *ideal depolarizer* $\mathbf{M}_{\Delta 0}$, the *3D depolarizer* $\mathbf{M}_3$ and the *2D depolarizer* $\mathbf{M}_2$. In fact, it is remarkable that the coefficient $(1 - P_3)$ of $\mathbf{M}_{\Delta 0}$ involves a contribution of amount $\hat{\lambda}_3$ from each of the four pure components of the spectral decomposition (note that $1 - P_3 = 4\hat{\lambda}_3$); the coefficient of $\mathbf{M}_3$ involves a contribution of amount $\hat{\lambda}_2 - \hat{\lambda}_3$ from each of the three more significant spectral components $\left[P_3 - P_2 = 3(\hat{\lambda}_2 - \hat{\lambda}_3)\right]$, and the coefficient of $\mathbf{M}_2$ involves a contribution of amount $\hat{\lambda}_1 - \hat{\lambda}_2$ from each of the two more significant spectral components $\left[P_2 - P_1 = 2(\hat{\lambda}_1 - \hat{\lambda}_2)\right]$. This redistribution of the relative weights of the components of the characteristic decomposition with respect to those of the spectral one, has consequences for the filtering processes considered in further sections.

To complete this section, it is worth recalling that, despite the fact that the detailed information of polarimetric purity is provided by the three IPP of $\mathbf{M}$, a proper overall measure of the polarimetric purity is given by the *degree of polarimetric purity* (or *depolarization index*) [26]

$$P_\Delta = \sqrt{\frac{\text{tr}(\mathbf{M}^T \mathbf{M}) - m_{00}^2}{3 m_{00}^2}} = \frac{1}{\sqrt{3}}\sqrt{2P_1^2 + \frac{2}{3}P_2^2 + \frac{1}{3}P_3^2} \qquad (20)$$

which is the 4D version of the n-dimensional degree of polarimetric purity associated with a n-dimensional covariance matrix $\boldsymbol{\Omega}$ (and, in particular associated with *n*-dimensional coherency and density matrices) defined as [20]

$$P_{nD} = \sqrt{\frac{1}{n-1}\left(\frac{n\,\text{tr}(\boldsymbol{\Omega}^2)}{(\text{tr}\,\boldsymbol{\Omega})^2} - 1\right)} = \sqrt{\frac{n}{n-1}\left(\sum_{k=1}^{n-1}\frac{P_k^2}{k(k+1)}\right)} \qquad (21)$$

where $P_k$ are the $n-1$ indices of polarimetric purity (IPP) of $\boldsymbol{\Omega}$.

Conversely, a proper measure of the overall polarimetric randomness associated with $\mathbf{M}$ is given by its depolarizance $D_\Delta$, defined as [27]

$$D_\Delta = \sqrt{\frac{4 m_{00}^2 - \text{tr}(\mathbf{M}^T \mathbf{M})}{2 m_{00}^2}} = \frac{\sqrt{9 - 6P_1^2 - 2P_2^2 - P_3^2}}{3} \qquad (22)$$

## 4. Covariance filtering

A commonly accepted first step of filtering a measured Mueller matrix $\mathbf{M}_{\exp}$ consists of obtaining an estimate of $\mathbf{M}_{\exp}$ that satisfies the covariance criterion through the following procedure:

1. Calculate $\mathbf{H}_{\exp} \equiv \mathbf{H}(\mathbf{M}_{\exp})$ through Eq. (2).

2. Calculate numerically the orthonormal eigenvectors of $\mathbf{H}_{\exp}$ as well as their eigenvalues $\lambda_{\exp i}$. Preferably, arrange the eigenvectors as the columns of the unitary matrix $\mathbf{U}$ that diagonalizes the Hermitian matrix $\mathbf{H}_{\exp}$ in such a manner that the $\lambda_{\exp i}$ appear in decreasing order $(\lambda_{\exp 0} \geq \lambda_{\exp 1} \geq \lambda_{\exp 2} \geq \lambda_{\exp 3})$ in the diagonalized version of $\mathbf{H}_{\exp}$.

3. Since negative values of $\lambda_{\exp i}$ are unphysical, and thus they are necessarily caused by experimental errors, replace all negative $\lambda_{\exp i}$ by zero. Hereafter we denote by $\lambda_{\text{cov} i}$ the resultant *covariance-filtered* eigenvalues obtained after the indicated replacements. Note that positive eigenvalues of the same order of magnitude as the predetermined tolerance $\delta$ may be replaced by zero, but, as indicated in Section 5, it would be preferable to keep them until the subsequent noise filtering step.

4. Build the covariance-filtered covariance matrix as

$$\mathbf{H}_{\text{cov}} = \mathbf{U}\,\text{diag}(\lambda_{\text{cov}0}, \lambda_{\text{cov}1}, \lambda_{\text{cov}2}, \lambda_{\text{cov}3})\mathbf{U}^{\dagger} \quad (23)$$

Obviously, it is expected that, at least, $\lambda_{\text{cov}0} = \lambda_0$, otherwise the measurement should be rejected because of its lack of quality.

5. Build the covariance-filtered Mueller matrix $\mathbf{M}_{\text{cov}} = \mathbf{M}(\mathbf{H}_{\text{cov}})$. The explicit general expression of $\mathbf{M}$ in terms of its associated covariance matrix $\mathbf{H}$ can be found in Ref. [28] (note that $\mathbf{H}$ is denoted as $\mathbf{C}$ in this and other references).

It should be stressed that, although the spectral decomposition is a particular option among the infinite possible arbitrary decompositions, it has the peculiarity that the weight $\hat{\lambda}_0$ of the pure component $\mathbf{M}_{U0}$ is maximal with respect to the coefficients of the pure components of all achievable arbitrary decompositions. In other words, $\hat{\lambda}_0$ represents the maximum achievable relative weight of a pure component with respect to all arbitrary decompositions of $\mathbf{M}_{\text{cov}}$, and hence justifies the use of the covariance filtering for the identification of the optimal pure estimate of a measured Mueller matrix. However, sometimes it occurs that it is expected (from the mere knowledge that the experimentalist has about the sample) that the representative pure component does not coincide with $\mathbf{M}_{U0}$; in which case the arbitrary decomposition and the polarimetric subtraction procedure [29], have to be considered and applied.

Furthermore, in the case of *blind measurements*, where the operator lacks completely of additional information on the nature or properties of the sample, the filtering process ends once the above covariance filtering has been applied to calculate $\mathbf{M}_{\text{cov}}$. Nevertheless, it is quite common that the experimentalist has certain information on the nature and structure of the sample, so that he has the founded expectation that the number $q$ of independent (incoherent) components mixed in the sample is less than 4. In the next section we analyze separately the cases for the different values of $q$.

## 5. Filtering of polarimetric noise

When the polarimetric randomness is not negligible and the interest is not only focused on the polarimetric nature of the characteristic component (determined by $\mathbf{M}_{U0}$), but also on its effective relative weight after a proper assessment of the noise structure, the characteristic decomposition has to be considered additionally for the filtering process.

Once the covariance filtering has been performed, $\mathbf{M}_{\text{cov}}$ can be submitted to its characteristic decomposition [see Eq. (15)], and, then, the values of the coefficients $P_1$, $P_2 - P_1$, $P_3 - P_2$, $1 - P_3$ of the respective components should be inspected in order to remove those whose coefficients have values close enough to the predetermined tolerance $\delta$. After this operation, the coefficients of the remaining components should be renormalized in order to satisfy that they sum to one.

It should be stressed that, since the spectral decomposition is a particular case among the infinite possible arbitrary decompositions, the procedure indicated in the above paragraph is clearly preferable (because of its objectivity) to the common criterion of removing the spectral components whose eigenvalues have the same order of magnitude as the predetermined tolerance $\delta$.

Let us now consider the Mueller matrix $\mathbf{M}_{\text{cov}}$ delivered by the covariance filtering process together with the number $q$ of effective parallel components hypothesized by the experimentalist in charge of the polarimetric analysis. Prior to tackle the possible situations, it is worth noting that, as stems from the mathematical formulation of the arbitrary decomposition, the maximum number of independent parallel components equals the rank $r$ of $\mathbf{H}_{\text{cov}}$ $(r \equiv \text{rank}\,\mathbf{H}_{\text{cov}} \leq 4)$, and therefore, when $q > r$, it is not possible to identify, or to decouple, the $q$ expected components from the measured Mueller matrix. Thus, strategies involving several measurements (as, for instance, for respective spatially separated parts of the sample) are necessarily required for the identification of more than $r$ components. Let us analyze next the cases where $q \leq 4$.

a $q = 4$. Since 4 is the maximum number of separable independent components, the covariance filtering, itself, completes the process. Obviously, if $r < 4$, then the hypothesis $q = 4$ does not match with the measurement and should be revised.

b $q = 3$. In this case, the filtering process, beyond the covariance one, depends on the value of *r*.

b.1 If $r = 4$, the contribution of the ideal depolarizer $\mathbf{M}_{\Delta 0}$ in the characteristic decomposition of $\mathbf{M}_{\text{cov}}$ should be considered as produced by a fully random noise. Note that a significant coefficient of $\mathbf{M}_{\Delta 0}$ $(1 - P_3 \gg \delta)$ implies necessarily the presence of four nonnegligible components, what is against the hypothesis that $q = 3$. Thus, the filtered Mueller matrix $\mathbf{M}_f$ is calculated as follows by subtracting $\mathbf{M}_{\Delta 0}$ from $\mathbf{M}_{\text{cov}}$

$$\mathbf{M}_f = \frac{\mathbf{M}_{\text{cov}} - (1 - P_3)\mathbf{M}_{\Delta 0}}{P_3}$$

$$= \frac{P_1 \mathbf{M}_{U0} + (P_2 - P_1)\mathbf{M}_2 + (P_3 - P_2)\mathbf{M}_3}{P_3} \quad (24)$$

and, after the appropriate renormalization (caused by neglecting the "white noise" associated with $\mathbf{M}_{\Delta 0}$), the spectral decomposition of $\mathbf{M}_f$ takes the form

$$\mathbf{M}_f = p_0 \mathbf{M}_{U0} + p_1 \mathbf{M}_{U1} + p_2 \mathbf{M}_{U2} \quad (25)$$

where the coefficients of the convex sum are given by

$$p_0 = \frac{\hat{\lambda}_0 - \hat{\lambda}_3}{1 - 4\hat{\lambda}_3} = \frac{1}{3}\left(1 + \frac{P_2}{2P_3} + \frac{3P_1}{2P_3}\right)$$

$$p_1 = \frac{\hat{\lambda}_1 - \hat{\lambda}_3}{1 - 4\hat{\lambda}_3} = \frac{1}{3}\left(1 + \frac{P_2}{2P_3} - \frac{3P_1}{2P_3}\right)$$

$$p_2 = \frac{\hat{\lambda}_2 - \hat{\lambda}_3}{1 - 4\hat{\lambda}_3} = \frac{1}{3}\left(1 - \frac{P_2}{P_3}\right) \quad (26)$$

Thus, after the above second step of the filtering process, the respective relative weights of the pure components $\mathbf{M}_{U0}$, $\mathbf{M}_{U1}$ and $\mathbf{M}_{U2}$ of $\mathbf{M}_f$ are not proportional to those of the spectral decomposition of $\mathbf{M}_{\mathrm{cov}}$, showing that, unlike the above formulation, the common procedure based on the mere elimination of the fourth component of the spectral decomposition of $\mathbf{M}_{\mathrm{cov}}$, followed by the subsequent renormalization of the coefficients, may lead to an incorrect evaluation of the coefficients of the components of the spectral decomposition of the filtered Mueller matrix.

b.2 If $r = 3$, then simply $\mathbf{M}_f = \mathbf{M}_{\mathrm{cov}}$.

c $q = 2$. In this case, the subsequent step after the covariance filtering depends on the value of *r*.

c.1 If $r > 2$, the contributions of $\mathbf{M}_{\Delta 0}$ and $\mathbf{M}_3$ in the characteristic decomposition of $\mathbf{M}_{\mathrm{cov}}$ should be considered as produced by a combination of a fully random noise (4D randomness) and the 3D random noise associated with $\mathbf{M}_3$. Thus, the filtered Mueller matrix $\mathbf{M}_f$ is calculated as follows through the appropriate subtraction of $\mathbf{M}_{\Delta 0}$ and $\mathbf{M}_3$ from $\mathbf{M}_{\mathrm{cov}}$

$$\mathbf{M}_f = \frac{\mathbf{M}_{\mathrm{cov}} - (1 - P_3)\mathbf{M}_{\Delta 0} - (P_3 - P_2)\mathbf{M}_3}{P_2}$$
$$= \frac{P_1 \mathbf{M}_{U0} + (P_2 - P_1)\mathbf{M}_2}{P_2} \quad (27)$$

and, after the appropriate renormalization (caused by neglecting the noise associated with the combination of $\mathbf{M}_{\Delta 0}$ and $\mathbf{M}_3$), the spectral decomposition of $\mathbf{M}_f$ takes the form

$$\mathbf{M}_f = p_0 \mathbf{M}_{U0} + p_1 \mathbf{M}_{U1} \quad (28)$$

where the coefficients of the convex sum are given by

$$p_0 = \frac{\hat{\lambda}_0 - \hat{\lambda}_2}{\hat{\lambda}_0 + \hat{\lambda}_1 - 2\hat{\lambda}_2} = \frac{1}{2}\left(1 + \frac{P_1}{P_2}\right)$$
$$p_1 = \frac{\hat{\lambda}_1 - \hat{\lambda}_2}{\hat{\lambda}_0 + \hat{\lambda}_1 - 2\hat{\lambda}_2} = \frac{1}{2}\left(1 - \frac{P_1}{P_2}\right) \quad (29)$$

As occurs in case (b.1), after the second step of the filtering process, the respective relative weights of the pure components $\mathbf{M}_{U0}$, and $\mathbf{M}_{U1}$ of $\mathbf{M}_f$ are not proportional to those of the spectral decomposition of $\mathbf{M}_{\mathrm{cov}}$.

c.2 If $r = 2$, then simply $\mathbf{M}_f = \mathbf{M}_{\mathrm{cov}}$.

d $q = 1$. This is the commonest case, where the interest is focused in the optimal pure estimate $\mathbf{M}_{U0}$ of $\mathbf{M}_f$, whose relative weight after subtracting the 4D, 3D and 2D contributions to the noise is given by $P_1$.

The Mueller matrix $\mathbf{M}_f$ obtained through the consecutive covariance and noise filtering procedures constitutes proper estimate for the Mueller matrix of the sample. Obviously, when appropriate, $\mathbf{M}_f$ can be submitted to further parallel or serial decompositions in order to obtain equivalent systems with the same polarimetric behavior as that corresponding to $\mathbf{M}_f$.

## 6. On subtracting the Mueller matrix of a predetermined constituent

Once the covariance and noise filtering processes have been performed, it may occur that the operator (due to its previous knowledge of the nature or characteristics of the sample) expects that a certain pure component, with Mueller matrix $\mathbf{M}_h$, should appear embedded in the sample as a parallel component (for instance, a known background or substrate component). In this case, the physical realizability of such hypothesis is checked through the following condition [29]

$$\mathrm{rank}(\mathbf{M}_f + \mathbf{M}_h) = \mathrm{rank}\,\mathbf{M}_f \quad (30)$$

so that the fulfillment of Eq. (30) allows for expressing $\mathbf{M}_f$ as the following convex combination of $\mathbf{M}_h$ and a remainder Mueller matrix $\mathbf{M}_x$

$$\mathbf{M}_f = p_h \mathbf{M}_h + (1 - p_h)\mathbf{M}_x \quad (31.\mathrm{a})$$

with [29]

$$p_h = \mathrm{tr}(\mathbf{\Lambda}_f \hat{\mathbf{H}}_h) \qquad \mathbf{M}_x = \frac{\mathbf{M}_f - p_h \mathbf{M}_h}{(1 - p_h)}$$
$$(\mathrm{rank}\,\mathbf{M}_f = \mathrm{rank}\,\mathbf{M}_h + \mathrm{rank}\,\mathbf{M}_x) \quad (31.\mathrm{b})$$

where $\mathbf{\Lambda}_f \equiv \mathrm{diag}(\hat{\lambda}_{f0}, \hat{\lambda}_{f1}, \hat{\lambda}_{f2}, \hat{\lambda}_{f3})$ is the diagonal matrix whose entries $\hat{\lambda}_{fi}$ are the ordered normalized eigenvalues of the covariance matrix $\mathbf{H}_f$ associated with $\mathbf{M}_f$ and $\hat{\mathbf{H}}_h \equiv \mathbf{H}_h/m_{00}$ is the normalized covariance matrix associated with the hypothesized constituent represented by $\mathbf{M}_h$.

Eq. (31) provides both the relative weight $p_h$ of $\mathbf{M}_h$ in the composed sample as well as both the relative weight $(1 - p_h)$ and the Mueller matrix $\mathbf{M}_x$ of the remainder unknown component.

## 7. Representative pure component

As seen in Section 2, the characteristic component $\mathbf{M}_{Uf0}$ of $\mathbf{M}_f$ has the largest relative weight with respect to any other arbitrary decomposition of $\mathbf{M}_f$ and therefore $\mathbf{M}_{Uf0}$ constitutes the optimal pure representative of $\mathbf{M}_f$ (and hence of $\mathbf{M}_{\mathrm{exp}}$). To go deeper into the interpretation of $\mathbf{M}_{Uf0}$, let us recall the symmetric decomposition of $\mathbf{M}_f$ [30]

$$\mathbf{M}_f = \mathbf{M}_{J2} \mathbf{M}_\Delta \mathbf{M}_{J1} \quad (32)$$

where $\mathbf{M}_{J1}$ and $\mathbf{M}_{J2}$ are pure Mueller matrices and $\mathbf{M}_\Delta$ is the canonical central depolarizer [31]. The characteristic decomposition of $\mathbf{M}_\Delta$ is given by the expansion

$$\mathbf{M}_\Delta = \mathcal{P}_1 \mathbf{M}_{C0} + (\mathcal{P}_2 - \mathcal{P}_1)\mathbf{M}_{C2}$$
$$+ (\mathcal{P}_3 - \mathcal{P}_2)\mathbf{M}_{C3} + (1 - \mathcal{P}_3)\mathbf{M}_{\Delta 0} \quad (33)$$

where $\mathcal{P}_i$ are the IPP of $\mathbf{M}_\Delta$, $\mathbf{M}_{C0}$ is the characteristic component, $\mathbf{M}_{C2}$ is the 2D depolarizer, $\mathbf{M}_{C3}$ is the 3D depolarizer and $\mathbf{M}_{\Delta 0}$ is the ideal depolarizer of the characteristic spectrum of $\mathbf{M}_\Delta$. In passing, it is worth recalling that, as shown in Ref. [32], $\mathbf{M}_{C0}$ is the identity matrix when $\mathbf{M}_f$ is type-I, and it corresponds to a diattenuator when $\mathbf{M}_f$ is type-II. It is remarkable that, since the number $\eta$ of 1-valued IPP of $\mathbf{M}_\Delta$ is equal to that of $\mathbf{M}_f$ [32], it turns out that the so-called *reference pure Mueller matrix* $\mathbf{M}_J(\mathbf{M}_f) \equiv \mathbf{M}_{J2} \mathbf{M}_{C0} \mathbf{M}_{J1}$ of $\mathbf{M}_f$ coincides with its characteristic component $\mathbf{M}_{Uf0}$ [32]

$$\mathbf{M}_J(\mathbf{M}_f) = \mathbf{M}_{Uf0} \quad (34)$$

This identity highlights the central role played by $\mathbf{M}_{Uf0}$ as the pure representative of $\mathbf{M}_f$.

## 8. Conclusion

Despite the fact that several procedures can be applied for the filtering of measured Mueller matrices in order to calculate their appropriate estimates [1-7], some general criteria have been analyzed from the point of view of the recent approaches given by the arbitrary and characteristic decompositions.

The covariance filtering transforms the measured Mueller matrix $\mathbf{M}_{exp}$ into a well-defined Mueller matrix $\mathbf{M}_{cov}$ and constitutes a first step for the determination of a proper global estimate of $\mathbf{M}_{exp}$. That is, the covariance filtering eliminates the part of the noise that is not compatible with the covariance criterion required for physically realizable Mueller matrices [1]. Thus, the remaining noise appears in the form of 4D, 3D or 2D polarimetric randomness that can be analyzed in an objective way by means of the characteristic decomposition of $\mathbf{M}_{cov}$, which provides the basis for a possible second step of noise filtering leading to the representative Mueller matrix $\mathbf{M}_f$. Depending on the previous knowledge of the nature of the sample, as well as on the particular aim of the measurement, $\mathbf{M}_f$ can be submitted to its arbitrary decomposition (including the commonly used spectral one) or to the polarimetric subtraction of known parallel components embedded in the sample. The physical realizability of the hypothesized presence of a predetermined pure component $\mathbf{M}_h$ within the medium represented by $\mathbf{M}_f$ can be checked by means of Eq. (30). In the absence of such hypothesis, the spectral decomposition

$$\mathbf{M}_f = \sum_{i=0}^{3} \hat{\lambda}_{fi} \left[ m_{00} \hat{\mathbf{M}}_{Ufi} \right]$$
(35)

constitutes an useful representation of the sample in the sense that the characteristic component $\mathbf{M}_{Uf0}$ (which coincides with the reference pure Mueller matrix of $\mathbf{M}_f$ [32]) is associated with the maximal relative weight $\hat{\lambda}_{f0}$, while the less significant spectral component $\mathbf{M}_{Uf3}$ is affected by the minimal relative weight $\hat{\lambda}_{f3}$ (with respect to all possible arbitrary decompositions of $\mathbf{M}_f$).

**Funding Information.** Ministerio de Economía y Competitividad (FIS2014-58303-P); Gobierno de Aragón (E99).